\documentstyle[12pt]{article}
\let\LARGE=\Large
\let\Large=\large
\let\large=\normalsize
\newcommand{\be}[3]{\begin{equation}  \label{#1#2#3}} 
\newcommand{\ee}{ \end{equation}}
\newcommand{\ba}{\begin{array}}
\newcommand{\ea}{\end{array}}

\newcommand{\x}{\times}
\newcommand{\ra}{\rightarrow}

\def\beq{\begin{equation}}
\def\eeq{\end{equation}}
\def\beqa{\begin{eqnarray}}
\def\eeqa{\end{eqnarray}}

\begin{document}
\thispagestyle{empty}
\rightline{IASSNS-HEP-98-17}
\rightline{HUB-EP-98/13}
\rightline{hep-th/9802193}
\vspace{2truecm}
\centerline{\LARGE {\bf
New $N=1$ Supersymmetric 3-dimensional Superstring}}

\vspace{0.3truecm}
\centerline{\LARGE \bf
Vacua from U-manifolds} 
\vspace{1.2truecm}
\centerline{Gottfried Curio$^1$ and
Dieter L\"ust$^2$\footnote{curio@ias.edu, 
luest@qft1.physik.hu-berlin.de\\
The first author is partially supported by NSF grant DMS 9627351}}
\vspace{.5truecm}
{\em 
\centerline{(1) School of Natural Sciences, Institute for Advanced Study,
 Princeton, NJ 08540}
\centerline{(2) Humboldt-Universit\"at, Institut f\"ur Physik, 
Invalidenstra\ss e 110, 10115 Berlin, Germany}}

\vspace{1.2truecm}
\vspace{.5truecm}
Making use of non-perturbative U-duality symmetries of type II strings
we construct new `superstring' vacua in three dimensions with N=1  
supersymmetry. This has an interpretation as compactifying formally 
from 13 dimensions (S-theory) on Calabi-Yau 5-folds possessing a
$T^3 \x T^2$ fibration. We describe some part of the massless multiplets,
given by the Hodge spectrum, and point to a corresponding 5-brane 
configuration.
\bigskip \bigskip
\newpage

\section{S-Theory}

The F-theory construction \cite{V,MV} can be generalized 
by considering the type IIB
string compactified (on a torus) to lower ($d'$) dimensions \cite{KuVa}. 
If one allows the
scalar fields (U-fields) to vary over some part of the $d'$-dimensional 
space and allows them to jump, consistent with the U-dualities, this data
will translate to a (complex) n-dimensional manifold $K^n$ whose (real)
$b$-dimensional base $B^b$ is the visible space from $d'$ to $d'-b$
dimensions and the (real) $2n-b$ dimensional fibre being the 
geometrization
of the U-duality. Hence this construction leads to  type IIB string
vacua with $d=d'-b$ flat, uncompactified dimensions.
As an example consider
S-theory \cite{KuVa,LuMin} with $d'=8$. In this case 
the scalar moduli space is given by the coset
\beqa
Sl(3,{\bf Z})\x Sl(2,{\bf Z})\backslash 
Sl(3,{\bf R})\x Sl(2,{\bf R})/SO(3)\x SO(2),\label{modulspace}
\eeqa
and
the U-duality group
is $Sl(3,{\bf Z})\times Sl(2,{\bf Z})$.  
In S-theory the seven scalar fields, which parametrize the above coset,
are allowed to vary over the 5-dimensional base $B^5$. 
The U-duality group
arises as a combination of two contributions: on the one hand one has
the $Sl(2,{\bf Z})$ which exists already
in 10 dimensions (the `S-duality' of the type IIB string), 
and is used there to
append the F-theory elliptic torus, leading to a theory living formally 
in 12 dimensions. This $Sl(2,{\bf Z})_S$ is united with the 
$Sl(2,{\bf Z})_T\times
Sl(2,{\bf Z})_U$ which arises on the other hand 
after compactification of the type IIB theory to 8 dimensions 
on a $T^2$; to be precise, $Sl(2,{\bf Z})_T$ 
combines with the $Sl(2,{\bf Z})_S$
to the $Sl(3,{\bf Z})$, whereas the $Sl(2,{\bf Z})_U$ remains giving
the $Sl(2,{\bf Z})$ factor of $Sl(3,{\bf Z})\x Sl(2,{\bf Z})$. Note 
that after compactification on the $T^2$ to 8 dimensions the theory 
becomes equivalent to type IIA and thereby to $M$ theory on $T^3$, 
which gives a further view on the U-duality group $Sl(3,{\bf Z})\x
Sl(2,{\bf Z})$.
So the (possibly reducible) Calabi-Yau
5-fold $K^5$ must be a $T^3\times T^2$
fibration over $B^{5}$:
\beqa
K^5\rightarrow_{T^3\times T^2}B^{5}.\label{stheory}
\eeqa
Since one has appended in 8 dimensions a 5-dimensional torus $T^3\times 
T^2$, $S$-theory can be regarded as a 13-dimensional theory.

Next we discuss what kind of 5-folds $K^5$ with $T^3\times T^2$ 
fibration can
be constructed. 
First consider {\em splitting}
(reducible) 5-folds which lead to $N=2$ supersymmetry in
3 dimensions.
One possible choice \cite{KuVa}
is a product space where one factor is a Calabi-Yau 3-fold $CY^3$ 
with $T^3$ fibration and the other factor is an elliptic $K3$, i.e.
\beqa
K^5=CY^3\times K3,
\qquad CY^3\rightarrow_{T^3}B^3\; , \, K3\ra _{T^2}S^2.\label{smanifold}
\eeqa
Therefore the total 5-dimensional base $B^5$ is given by
\beqa
B^5=B^3\times S^2.\label{tbase}
\eeqa
We will always assume that $B^3=S^3$ (cf. also \cite{LuMin}; this is
connected with mirror symmetry on Calabi-Yau 3-folds \cite{SYZ},
\cite{GW}).
Another class of vacua with $N=2$ supersymmetry is given by
\beqa
K^5=CY^4\times T^2,\qquad CY^4\rightarrow_{T^3}B^5.
\eeqa
The $CY^4$ is assumed to be $CY^3$ fibered over $P^1$.
Therefore $B^5$ is a 
$S^3$ fibration over $S^2$.

In this paper we will see (section 3) that Calabi-Yau 5-folds exist 
which are $T^3\times T^2$ fibrations but not product spaces. 
This leads to $N=1$ supersymmetry in 3 dimensions. 
Before, this was, in string compactification, possible to be reached 
only using the somewhat difficult to handle
spaces of exceptional holonomy (heterotic string on $G_2$-manifold,
M-theory on $Spin(7)$-manifold). By contrast it is realised here still
in the framework of the well-suited Calabi-Yau spaces. 

The new three-dimensional superstring vacua described in this paper 
might also lead to a geometric understanding of (non-perturbative) 
effects in $3D$, $N=1$ supersymmetric field theory  \cite{AHW}, 
possibly very much like the realisation of instantons, contributing 
to the superpotential, via internal geometrical cycles in the context 
of $M$-theory resp. $F$-theory on a Calabi-Yau four-fold 
(leading to $N=2$ in $3D$ resp. $N=1$ in $4D$).

Finally in view of Witten's szenario \cite{wit2} relating $N=1$ 
supersymmetric theories in three dimensions to non-supersymmetric 
theories in four dimensions with vanishing cosmological constant
one can try to relate the described 3D vacua to 4D vacua of $N=0$
like it was tried \cite{V} for $M$-theory on $Spin(7)$ manifolds.

\section{Some dualities}

Let us first recall \cite{KuVa} the duality symmetries between 
S-theory on the
one hand and F-theory and the heterotic string on the other hand.
One derives first  that S-theory on $K^5\times S^1$ is 
dual to F-theory on $K^5$:
\beqa
d=2:\quad S|_{K^5\times S^1}\quad \leftrightarrow\quad F|_{K^5},\label{sf}
\eeqa
Here in $F|_{K^5}$ - i.e. as soon as one has leaved S-theory, 
which is in a
sense a type IIB theory with additional structure, and has reached 
$F$-theory, which is a type IIB theory with a different additional
structure - the $T^2$ fibre of $K^5$ corresponds to the 
elliptic fibre used in F-theory to codify the type IIB 
complex coupling constant. 
On the other hand, the volume of the $T^3$ fibre of $K^5$ in F-theory 
corresponds to the inverse radius of $S^1$.
One can go further down in dimensions and arrives at the following chain 
of dualities
\beqa
d=1:\quad S|_{K^5\times T^2}\quad \leftrightarrow\quad F|_{K^5\times S^1}
\quad \leftrightarrow \quad M|_{K^5},\label{fm}
\eeqa
where the inverse radius of the extra circle is related 
to the volume of the $T^2$ fibre in M-theory. 

Before we discuss S-theory on $CY^5$ in greater detail let us consider the
reducible cases of S-theory on $K^5=CY^3\times K3$ with an elliptically
fibered $K3$ resp. on $K^5=CY^4\times T^2$ with a $CY^3$ fibered $CY^4$.
In case of S-theory on $K^5=CY^3\times K3$
the above chain of dualities can be extended \cite{KuVa}. As,
upon compactification on $S^1$, this is dual to F-theory on $CY^3\times K3$ and
F-theory on the elliptic $K3$ is dual
%\footnote{Note also that if $CY^3$ is $K3$ fibered one gets a duality with 
%F-theory on $CY^4\x T^2$ (i.e. type IIA on $CY^4$): compare \cite{BeSa} where 
%F-theory on $K3\x K3$ is related to F-theory on $CY^{43,43}\x T^2$, extend 
%this relation in its first factors then adiabatically over a further $P^1$.}
to the heterotic string on $T^2$,
this means that in 2 dimensions we have a duality with a heterotic 
string on $CY^3\times T^2$:
\beqa
d=2:\quad S|_{CY^3\times K3\times S^1}
\quad \leftrightarrow\quad H|_{CY^3\times T^2},\label{ds22}
\eeqa
The other reducible case of S-theory on $K^5=CY^4\times T^2$ with a $CY^3$ 
fibered $CY^4$ is also interesting to consider and {\em not} already 
covered (via duality) by some other theory known before, especially it 
{\em is not} dual to $M$-theory on $CY^4$. For this note that after 
compactification on $S^1$ the $T^2$ factor is now the $F$-theory elliptic 
fibre, i.e. this is simply type IIB on $CY^4$ of $(0,4)$ spacetime 
supersymmetry in 2 dimensions. By
contrast $M$-theory on $CY^4$ is the lifting to 3 dimensions 
of the non-chiral theory
of $(2,2)$ space-time supersymmetry in 2
dimensions given by type IIA on $CY^4$; in other words
the 13-dimensional S-theory on $CY^4\x T^2$ can be viewed as the lift
of type IIB
%there is a seemingly 11D theory $M'$ (which is actually the 13D theory S on
%$CY^4\x T^2$) which lifts type IIB 
on a $CY^4$ from 2 to 3 dimensions just as the
11-dimensional $M$-theory does the corresponding thing for type IIA.

Let us come now to the 3-dimensional
 theories with $N=1$ supersymmetry like S-theory on $CY^5$ or
$M$-theory on a $Spin(7)$ manifold (or the heterotic string on a $G_2$ 
manifold). Let us consider, in view of the observation just made for the 
reducible $CY^4\x T^2$ case, again first the situation in 2
dimensions. There
one has again the non-chiral theory 
%of space-time  $(1,1)$ supersymmetry 
given by
type IIA on a $Spin(7)$ 
manifold and the chiral one 
%of space-time  $(0,2)$ supersymmetry 
given by
F-theory on $CY^5$.
%
%
% 
%Note that $F$-theory on $CY^5$ is OF WHICH OF THESE
%TYPES? 
Then $M$ theory on $Spin(7)$ lifts the first, non-chiral, theory to 3
dimensions,
whereas S-theory on $CY^5$ lifts the chiral theory to 3 dimensions.

%THE ISSUE OF FURTHER LIFTINGS TO 4D (SZENARIO) BY F ON $Spin(7)$ RESP.
%$CY^4/Z_2$?-- MAY BE MUSIC FOR THE FUTURE, IF NOT SILENCE.

\section{Some Calabi-Yau 5-folds}

For the compactification of S-theory we will now construct (complex) 
Calabi-Yau 5-folds 
which have a $T^3\times T^2$ fibration. The $T^3$ part we will always get
from a $CY^3$ (which is assumed to have a mirror, and so a $T^3$ 
fibration over a real 3-dimensional base $B^3=S^3$, cf. \cite{SYZ}). 
%So the 
%$following two possibilities for a {\em splitting} (reducible) 
%5-fold immediately
%come to mind: the $CY^3\times K3$ and the $CY^4\times T^2$, where the 
%$CY^4$ is assumed to be $CY^3$ fibered over $P^1$. The real 5-dimensional
%base spaces are respectively $S^3\x S^2$ and $B^5$, where $B^5$ is a 
%$S^3$ fibration over $S^2$.
We concentrate on
%Much more interesting are 
spaces which are true 
(irreducible, {\em non-splitting}) $CY^5$.
We will discuss below
examples of the form $X^4\x _{P^1}dP_9$, where $X^4$ is a 
(non Calabi-Yau) four-fold which 
has a $CY^3$ fibration over $P^1$ (whose fibration structure\footnote{For 
example one can have a description 
derived from a representation $CY^5=(CY^4\x T^2)/{\bf Z}_2$
which is to be understood in the same sense as the construction of the 
$CY^{19,19}=dP_9\x _{P^1} dP_9$ from the quotient $(K3\x T^2)/{\bf Z}_2$ 
(cf. appendix and \cite{CL}); similarly one could study a
$CY^5=(CY^3\x K3)/{\bf Z}_2$ version, this time with the 
$dP_9$-fibrefactor appearing by `reduction' from $K3$.}, which is also 
part of the `input data' structure and not given by the $CY^3$ alone, 
will be described below) and 
$dP_9:={\tiny \left[\begin{array}{c|c}P^2&3\\P^1&1\end{array}\right]}$
is a surface\footnote{cf. the appendix and \cite{DGW,CL,G}}
having an elliptic fibration over $P^1$.
Clearly, the very idea of this fibre product is that the splitting of
the fibration does not imply the splitting of the total space.
The real 5-dimensional base $B^5$ for the 
$T^3\times T^2$ fibration is  a $S^3$ fibration over $S^2$, which
will be described in more detail in the next section. 

For example one can build out of the Calabi-Yau 3-fold
$CY^{19,19}$, by fibering it over a further $P^1$, the 4-fold
$X^4={\tiny \left[\begin{array}{c|cc}
P^2&3&0\\P^1&1&1\\P^2&0&3\\P^1&0&1\end{array}\right]}=dP_9\x _{P^1} K$ 
with $K:={\tiny \left[\begin{array}{c|c}
P^1&1\\P^2&3\\P^1&1\end{array}\right]}$ and finally the 
Calabi-Yau 5-fold 
$CY^5={\tiny \left[\begin{array}{c|ccc}
P^2_a&3&0&0\\P^1_b&1&1&0\\P^2_c&0&3&0\\P^1_d&0&1&1\\P^2_e&0&0&3
\end{array}\right]}
=dP_9\x _{P^1}K\x_{P^1}dP_9$. We will analyse that example\footnote{Note that 
one has actually an
$T^2_a\x T^2_c\x T^2_e$ fibration over $P^1_b\x P^1_d$;
in the quotient description here the 
$CY^4=dP_9\x _{P^1} {\cal B}$ (cf. appendix) with
$E_8$ superpotential of \cite{DGW,CL} occurs.} 
further below.

 Let us analyse now the cohomology of a general\footnote{assumed to be 
non-splitting; 
the Hodge diamonds for the reducible cases $CY^4\x T^2$ and 
$CY^3\x K3$ are of course trivially computed.}
%for the convenience of the reader displayed in 
%the appendix.} 
(i.e. at first not necessarily of the fibre-product form) 
Calabi-Yau 5-fold. Its Hodge diamond looks like 
\beqa
{\scriptsize
\begin{array}{ccccccccccc}
 & & & & &1& & & & & \\
 & & & &0& &0& & & & \\
 & & &0& &h^{11}& &0& & & \\
 & &0& &h^{21}& &h^{21} & &0& & \\
 &0& &h^{31}& &h^{22}& &h^{31} & &0& \\
1& &h^{41}& &h^{32}& &h^{32} & &h^{41} & &1
\end{array}
}\nonumber
\eeqa 
 
 Below we will compute the Hodge numbers which have the most immediate 
interpretation, $h^{11}$ as K\"ahler parameters and $h^{41}$ as complex 
deformations, from our input data in the class of 
$X^4 \x _{P^1} dP_9$ spaces.
Furthermore from the $CY^3\x T^2$ fibration over $P^1$ of these spaces
one gets for the Euler number $e$ of the 5-fold $e=12\cdot e(CY^3)$ as 
$e(dP_9)=12$. So we will `know' 3 of the 6 unknown numbers. 
In the Calabi-Yau
4-fold case one gets one further information from a relation derived in
\cite{SVW}; this is enough for the 4 unknowns in the 4-fold case, in our 
case 2 unknowns remain. Let us see in detail how this happens.

The index of the (1,0)-forms-valued
$\bar{\partial}$ operator, $ind \, \bar{\partial}=
\sum_{q=0}^5 (-1)^q h^{q,1}$, is according to the index theorem
given by
\beqa
ind \, \bar{\partial}=\int_X Td(X)ch(T^* X),\nonumber
\eeqa
where for the Calabi-Yau 5-fold $X$ one has 
\beqa
Td(X^5_{CY})&=&1+\frac{c_2}{12}+\frac{3c_2^2-c^4}{720},\nonumber\\
ch(T^* X^5_{CY})&=&5-c_2+\frac{-c_3}{2}+\frac{c_2^2-2c_4}{12}+
\frac{-c_5-c_2\cdot (-c_3)}{24}\nonumber
\eeqa
so that one gets finally the relation
\beqa
-\frac{e}{24}=h^{41}-h^{31}+h^{21}-h^{11}.
\eeqa
Taken together with the obvious relation
\beqa
\frac{e}{2}=h^{11}-h^{41}+2(h^{31}-h^{21})+h^{22}-h^{32}
\eeqa
one can now express the cohomology completely in terms
of the known numbers $e$, $\delta:=h^{41}-h^{11}$ and the remaining 
unknowns $h^{21}$ and $h^{22}$ 
\beqa
h^{31}&=&h^{21}+\delta + \frac{e}{24}\nonumber\\
h^{32}&=&h^{22}+\delta - \frac{5}{12}e
\eeqa
%\beqa
%{\scriptsize
%\begin{array}{ccccccccccc}
% & & & & &1& & & & & \\
% & & & &0& &0& & & & \\
% & & &0& &h^{11}& &\, 0& & & \\
% & &0& &h^{21}& &h^{21} & &0& & \\
% &0\,\,& &h^{21}+\delta+\frac{e}{24}& &h^{22}& &h^{21}+\delta+\frac{e}{24}
%   & &0& \\
%1 & &h^{41}& &h^{22}+\delta -
%                      \frac{5}{12}e& &h^{22}+\delta-
%                     \frac{5}{12}e& & h^{41}& &1
%\end{array}
%}\nonumber
%\eeqa 
 
Also one has now for the 5-folds of the special form 
$CY^5=X^4\x _{P^1}dP_9$
that (cf. appendix; $CY^3$ denotes the fibre of $X^4$)
\beqa
h^{21}&=&h^{21}(X^4),\nonumber\\
h^{22}&=&10h^{11}(CY^3)+2h^{21}(X^4)+h^{22}(X^4)+1.
\eeqa

Now let us come back to our example $CY^5=dP_9\x _{P^1}K\x _{P^1}dP_9$ 
of $e=0$. This decomposition allows one to find for the K\"ahler 
classes $h^{11}=10-1+3+10-1=21$.
On the other hand one has (with $\# def dP_9=8$) 
for the complex deformations that
$h^{41}=8+3+\# def K +8+3$, so with
$\# def K=2\cdot 10\cdot 2-(3+8+3)-1=25$ one gets $h^{41}=47$.
Furthermore the decomposition shows (cf. the appendix) 
that $h^{21}=h^{21}(K)=\# def K-(\# def dP_9 -1)=
2\cdot 10\cdot 2-(3+8+3)-1-(8-1)=18$
and $h^{22}=10\cdot 19+2\cdot 18+h^{22}(X^4)+1$ which gives with
$h^{22}(X^4)=10\cdot 10+2h^{21}(X^4)+h^{11}(K)+1=140$ that $h^{22}=367$, 
so
\beqa
{\scriptsize
\begin{array}{ccccccccccc}
 & & & & &1& & & & & \\
 & & & &0& &0& & & & \\
 & & &0& &21& &0& & & \\
 & &0& &18& & 18& &0& & \\
 &0& &44& &367& & 44& &0& \\
1& &47& &393& & 393& & 47& &1
\end{array}
}\nonumber
\eeqa

Similarly one can use any of the existing lists of Calabi-Yau 4-folds
(for example with the $STU$-Calabi-Yau $P_{1,1,2,8,12}(24)$ as 3-fold
fibre), go to the correspondingly reduced $X^4$ (model $X^4_{A^{\prime}}$
for the example just mentioned, cf. \cite{CL}) and describe a $CY^5$.

\section{The brane point of view}

Just as one can interpret an F-theory vacuum either as a Calabi-Yau 
compactification of a formally twelfe-dimensional theory or as a 
type IIB vacuum with varying dilaton and 7-branes one can use the 
alternative brane point of view for S-theory as well. This was studied 
especially for the $T^3$ part in \cite{LuMin}; let us recall this point 
of view first and then interpret our example that way.

 So what is given according to this point of view is really an 8D vacuum
configuration with varying moduli consistent with the U-duality group
$Sl(3,{\bf Z})\x Sl(2,{\bf Z})$. The relevant moduli space 
eq.(\ref{modulspace})
%\beqa
%Sl(2,{\bf Z})\x Sl(3,{\bf Z})\backslash 
%Sl(2,{\bf R})\x Sl(3,{\bf R})/SO(2)\x SO(3)\label{modulspace}
%\eeqa
leads as described to the idea of U-manifolds admitting a $T^3\x T^2$ 
fibration. $T^2$ fibrations were studied in F-theory so let us focus 
on the fivedimensional piece of the moduli space parametrizing a 
three-torus of constant volume. This translates to a family of 5-branes
that transform consistently with $Sl(3,{\bf Z})$, living on $S^3$, where
each individual member lives on a `line' (set of real codimension two)
in the base $S^3$. 
(In addition the 5-branes are also
wrapped around the $S^2$ part
of the $B^5$ base.)
Phrased differently (and making it comparable to the 
cosmic string viewpoint in F-theory) the solution may be viewed as a 
mapping of the base into the five-dimensional moduli space, which is 
locally $Sl(3,{\bf R})/SO(3)$ and actually an orbifold from the 
identification under the action of $Sl(3,{\bf Z})$ U-duality. The pullback
of the orbifold singularities leads then to the 5-brane configuration
wrapping the singular lines (compare the F-theory picture relating
the degenerate elliptic curves with the 7-brane locations; here the 
$T^3$'s are expected to degenerate along the singular lines, which 
correspond to the one-dimensional
compact part of the world volume
 of the 5-branes; but note that in the F-theoretic 
case with the $K3$ there are only parallel branes involved).

Let us describe the real 5-dimensional base space $B^5$ for the 
$T^3\x T^2$ fibration of $CY^5=X^4\x _{P^1}dP_9$ more fully. It is 
as already
remarked a $S^3$ fibration over $S^2$. This simply connected space has
the one non-trivial Betti number $b_2=1(=b_3)$. Because of 
$\pi _1(SO(4))={\bf Z}_2$ there exist actually only two possibilities for 
$B^5$, either the product $S^2\x S^3$ or the twisted version. 
Furthermore, taking into account the possibility to represent the 
$S^3$ fibre itself as a $S^1$ fibration over $S^2_f$ (the Hopf 
fibration), the mentioned ambiguity for the $B^5$ space can, 
because of $\pi _1(SO(3))={\bf Z}_2$, already be 
read of from the 4-manifold consisiting of the $S^2_f$ fibration over 
the base $S^2$. Note that, in the case of a complex structure
for this 4-manifold, the `fibre-type' ($\in {\bf Z}_2$) ambiguity of the
Hirzebruch surface $F_n$ (being a $P^1_f=S^2_f$ fibration over the base
$P^1=S^2$) is described by $n$ being even/odd (note the deformation
$F_2\ra F_0$).

The $T^3$ fibration over $S^3$ of the $CY^3$ fibre
will be combined with the $T^2_e$ fibration over $P^1_d$ of the 
$dP_9^{de}$. The singular lines in the $S^3$ are now replaced by 
singular loci of real codimension two in the base $B^5$ (which itself
is a $S^3$ fibration over $S^2_d$).
Note that we get a 5-brane picture
in total as the $T^2$ fibration part
%relates to 
%(not in 8D but at least after going to $F$-theory, 
%via reduction on a $S^1$, 
%is U-conjugated to 
%the $Sl(2,{\bf Z})$ part of the U-duality group, 
%which exists already
%in 10D (S-duality) and is there codified by F-theory, 
gives also 5-branes:
%(after again decompactifying the auxiliary $S^1$): 
namely 7-branes (cf. F-theory) compactified
on the $T^2$ which brought us from 10 to 8 dimensions; note that here
the 7-branes have their locus not on a $P^1$, compactifying 10 
dimensions to 8 dimensions,
but on the $P^1_d$, compactifying from 5 dimensions to 3 dimensions.
Of course the relevant singular loci
consist now in the singular lines of the three-fold with their
parameters running over the 
$S^2_b$ base of $B^5$ on the one hand, and furthermore in the $S^3$
fibers over the 12 singular points (for the $dP_9^{de}$) on $S^2_d$.
So one gets in both cases subspaces of real dimension three in $B^5$,
i.e. the loci of the 5-branes are of real codimension 2 in the base. 

Finally let us also consider the analogue of the (now not 
internal but spacetime-filling) 3-branes which have to be turned on for
F-theory on a 4-fold. These are, as already mentioned in \cite{KuVa},
(spacetime-filling) 4-branes in the case of S-theory on a 4-fold. Now
our $CY^5$ is $CY^4$ fibered over $P^1$; so in this further 
compactification process the 4-branes wrap the $P^1$ and become 
(spacetime-filling) membranes in three dimensions. But as our
$CY^4$ fibre had to be the reducible $CY^3\x T^2$ of Euler number zero,
the mentioned branes do not actually occur.

\section{The Spectrum}

Let us now read of from the Hodge diamond some part of 
the spectrum of massless multiplets. We will do this by the same 
strategy which is used to get a corresponding part of the F-theory 
spectrum from information about type II compactifications 
\cite{MV,CL}. This uses that F-theory on $X\x T^2$ is type IIA on $X$
and furthermore that F-theory, being partly simply type IIB on the 
basis $B$ of the relevant Calabi-Yau space in question, shows a 
sensitivity on the Hodge numbers of $B$ (which is not seen any longer
- after further compactification on $T^2$ - in the type IIA description).
Now in our case here we will use the same procedure and relate S-theory 
after further compactification on $T^2$ to M-theory. Two special features
appear in our setup: first in three dimensions both the possible 
multiplets, the scalars and the vectors, are actually, because of 
duality, in some sense indistinguishable; secondly, as we followed quite
strictly the adiabatic strategy (splicing a $CY^3$ and a $T^2$ together
over the new $P^1$), we are actually in a case which for F-theory on 
Calabi-Yau 3-folds would correspond to having only Hirzebruch 
surfaces as bases and no further birational transformations (blowings up
and down) made on the base (i.e. no non-trivial tensor multiplets); this
is reflected here in the property of $B_5$ being a $S^3$ fibration over 
$S^2$, so its cohomological data relevant here (Betti numbers) are fixed 
(cf. sect. 2).

Now, the searched for spectrum is that of S-theory on $CY^5$.
This leads to $N=1$ supersymmetry in three dimensions.
Besided the $N=1$ supergravity multiplet, which contains as its
bosonic degree of freedom the metric $g_{\mu\nu}$ ($\mu,\nu=0,1,2$),
there will be $S_3$ real $N=1$ scalar multiplets plus $V_3$ real
$N=1$ vector multiplets. The on-shell degrees of freedom of each $N=1$
scalar multiplet are given by one real scalar field  plus one real 
Majorana spinor; the $N=1$ vector multiplets in three dimensions contain,
on-shell, one vector field plus one real Majorana spinor. 
Since in three dimensions
a vector is dual to a scalar, there is a (supersymmetric) Poincare duality
between the scalar and the vector multiplets. 

To obtain the spectrum of $S$-theory on $CY^5$ we start with the 
consideration of the type IIB
superstring in ten dimensions. Its massless bosonic fields are
\beqa
g_{MN},~ B_{MN},~ \phi ,~\phi ', ~ A_{MN}, ~A^+_{MNPQ} .\label{IIB10}
\eeqa
To obtain $S$-theory vacua we first have to compactify the type IIB 
superstring
on a two-dimensional torus $T^2$ to eight dimensions. This leads to
non-chiral eight-dimensional $N=2$ supergravity
(like the type IIA compactification on
$T^2$). The only massless supermultiplet
is the supergravity multiplet. From eq.(\ref{IIB10}) it is easy to see
that the eight-dimensional $N=2$ supergravity 
multiplet contains the following massless bosonic fields:
\beqa
g_{MN},~ 7\phi,~ 6A_{M}, ~ 3 A_{MN}, ~A_{MNP}.
\label{IIB8}
\eeqa
(Now the indices $M,N,\dots$ run over $0,\dots ,7$; note that a possible
4-form is dual to the 2-form coming from the 10D 4-form which (i.e. its
field-strengh) is self-dual.)
The seven scalar fields parametrize the non-compact coset space 
eq.(\ref{modulspace}).
%and the $U$-duality group is  given by the discrete group
%$Sl(3,{\bf Z})\times Sl(2,{\bf Z})$.

At the next step we compactify this eight-dimensional theory
down to three dimensions on the $S$-theory base space $B^5$ 
to obtain the base-sensitive part of the three-dimensional spectrum. One 
performs the harmonic analysis on $B^5$ deriving from eq.(\ref{IIB8})
the following 
contributions $s_3$ and $v_3$  to the number of scalar and $U(1)$ vector
fields (as $b_1=0$, $b_2=b_3=1$):
\beqa
s_3=7+6b_1+3b_2+1b_3=11,\quad v_3=6+3b_1+b_2=7.\label{spectrs3}
\eeqa

Now we consider $M$-theory on $CY^5$, which leads to $N=2$
supergravity in one dimension. There are $V_1^M$  
vector multiplets with one real physical scalar and one non-propagating 
vector (plus one non-propagating scalar). In addition we will have 
$S_1^M$ scalar multiplets with each one  physical scalar 
field. The internal metric of $CY^5$  provides 
$h^{1,1}+2h^{4,1}$ real scalars. The 11-dimensional field $A_{MNP}$ 
will contribute in addition $h^{1,1}$ $U(1)$ vectors plus 
$2h^{2,1}$ scalars.
So in total we derive:
\beqa
V_1^M=h^{1,1},\quad S_1^M=2h^{2,1}+2h^{4,1}.\label{spectr1}
\eeqa
This $M$-theory spectrum can be directly compared with the $S$-theory 
spectrum on $CY^5\times T^2$. The three-dimensional massless 
spectrum of $S$-theory, denoted
by $S_3$ and $V_3$ is related to the one dimensional $S$-theory 
spectrum as follows:
\beqa
V_1^S=V_3+2,\quad S_1^S=S_3.
\eeqa
So with eq.(\ref{spectr1}) the $S$-theory/M-theory
duality in one dimension leads to  the constraint 
$V_1^S+S_1^S=V_1^M+S_1^M$,
and hence we derive
\beqa
V_3+S_3=h^{1,1}+2h^{2,1}+2h^{4,1}-2.
\eeqa
Including the base-sensitive part one gets
\beqa
S_3&=&2h^{2,1}+2h^{4,1}+s_3-v_3=2h^{2,1}+2h^{4,1}+4,\nonumber\\
V_3&=&h^{1,1}+v_3-s_3-2=h^{1,1}-6.
\eeqa

Note that the $V_3$ only counts the massless
abelian vector fields; at special loci
in the moduli space additional non-abelian gauge bosons together with 
charged matter fields are expected to become massless.

\smallskip
\noindent
We like to thank A. Miemiec, R. Minasian, C. Vafa and E. Witten
for useful discussions.

\appendix

\section{Appendix}

{\bf 5-folds as fibre products}

For more datails on many of the mentioned spaces cf. for example 
\cite{CL,DGW}.

The {\bf surface} $dP_9={\tiny 
\left[\begin{array}{c|c}P^2&3\\P^1&1\end{array}\right]}$ has the one 
nontrivial Hodge number $h^{11}=10$ and $8=10\cdot 2-(8+3)-1$ 
complex deformations. Note that you can visualize the 10 classes on the 
one hand by viewing the surface $P^2$ blown up in the 9 intersection 
points of the two cubics (in this sense it is a generalization of 
the del Pezzo surfaces $dP_i$ for $i=1,\dots , 8$); on
the other hand you can understand their appearence topologically
in the elliptic fibration picture via the fact that an $S^1$ of the 
fibre moving between two vanishing points traces out an $S^2=P^1$ 
(cf. \cite{G}).

The Calabi-Yau {\bf three-fold}
$CY^{19,19}={\tiny 
\left[\begin{array}{c|cc}P^2&3&0\\P^1&1&1\\P^2&0&3\end{array}\right]}
=dP_9\x _{P^1} dP_9$ has obviously $h^{11}=10+10-1$ and so from $e=0$ also
$h^{21}=19$, which you can also count directly as $8+3+8$ (as one can 
use the reparametrization freedom on the $P^1$ only once). Note that also
$CY^{19,19}=(K3\x T^2)/{\bf Z}_2$ (here the first 
of the two $dP_9$-fibre factors is appearing
`by reduction' from the former $K3=
{\tiny \left[\begin{array}{c|c} P^2&3\\P^1&2\end{array}\right]}$-factor, 
the second one is `emerging' from
the constant $T^2$-factor in the process of smoothing out the quotient).

The Calabi-Yau {\bf four-fold} $CY^4={\tiny 
\left[\begin{array}{c|cc}P^2&3&0\\P^1&1&1\\P^1&0&2\\P^2&0&3
\end{array}\right]}=dP_9\x _{P^1} {\cal B}$ with 
${\cal B}:={\tiny \left[\begin{array}{c|c}
P^1&1\\P^1&2\\P^2&3\end{array}\right]}$ 
has the following Hodge diamond (model A, cf. \cite{CL})
\beqa
{\scriptsize
\begin{array}{ccccccccccccccccccc}
{\Large {\bf A}}& & & &1& & & & &\qquad \qquad&{\LARGE {\bf A^{\prime}}}& & & 
&1& & & &  \\ 
  & & &0& &0& & & &\qquad \qquad&  & & &0& &0& & & \\
  & &0& &12& &0& & &\qquad \qquad&  & &0& &12& &0& &  \\ 
  &0& &28& &28& &0& &\qquad \qquad&  &0& &112& & 112& &0&  \\
 1& &56& &260& &56& &1&\qquad \qquad& 1& &140& &428& & 140& &1
\end{array}
}\nonumber
\eeqa 
where also $h^{21}=h^{21}({\cal B})$ (cf. for this numerically 
\cite{CL}; it can be seen 
also from the topological `tracing out' argument above that no 
new classes appear:
just as the $K3$ fibration of ${\cal B}$ over $P^1$ shows how a $S^2$ 
moving over the base between two vanishing points traces out a $S^3$ for
$h^{21}({\cal B})$, the corresponding $T^2\x K3$ fibration over $P^1$
of $CY^4=dP_9\x _{P^1} {\cal B}$ shows that no new classes appear). 
Note also that $CY^4=(T^2\x CY^3)/{\bf Z}_2$.
There is an `alternative version' of 
this space which uses instead of the $CY^3={\tiny \left[\begin{array}{c|c}
P^1&2\\P^1&2\\P^2&3\end{array}\right]}$, from which ${\cal B}$
is derived (by quadratic base change in the one $P^1$), 
the well-known $CY^{3,243}$; this leads to a $CY^4$ 
with Hodge-diamond shown above as model $\mbox{A}^{\prime}$ \cite{CL}.

So one has in both cases
\beqa
h^{22}=204+2h^{21}.
\eeqa
This relation which of course is not accidental
has two explanations: a numerical one and a geometrical one. The latter 
will be of relevance for our understanding of $h^{22}$ of a $CY^5$.

Now first the numerical argument: one has
\beqa
h^{22}=e-4-2(h^{11}-2h^{21}+h^{31})=\frac{2}{3}e+12+2h^{21}
\eeqa
(using $h^{11}-h^{21}+h^{31}=\frac{e}{6}-8$ (cf. \cite{SVW}))
and $e=12\cdot 24=288$ shows the relation asserted above. Secondly the 
geometrical interpretation makes visible the classes (of the relevant
Hodge type) from the following four topological sources of 4-cycles:\\ \\
$S^2\x S^2$: $20\cdot 10$ \\
$S^1\x $ 3-cycle: $2h^{21}$\\
4-cycle$\x$point: $h^{22}({\cal B})=h^{11}({\cal B})=3$\\
point$\x dP_9$: $1$.

So if we come now to the Calabi-Yau {\bf five-fold} 
$CY^5=X^4\x _{P^1}dP_9$ we have again $h^{21}=h^{21}(X^4)$ 
and in the case of $X^4=dP_9\x _{P^1}K$
further $h^{21}=h^{21}(K)$. Also we have from the geometrical arguments
showing how $h^{22}$ arises that (let $CY^3$ be the fibre of $X^4$ over 
$P^1$)
\beqa
h^{22}=10h^{11}(CY^3)+2h^{21}(X^4)+h^{22}(X^4)+1.
\eeqa

%Finally let us display the Hodge spectra of the reducible cases 
%(note that in the next two Hodge diamonds unspecified Hodge numbers 
%refer to the input $CY^4$ resp. $CY^3$).
%For $CY^4\x T^2$ one has (cf. M-theory on $CY^4$)
%\beqa
%{\scriptsize
%\begin{array}{ccccccccccc}
% & & & & &1& & & & & \\
% & & & &1& &1& & & & \\
% & & &0& &h^{11}+1& &0& & & \\
% & &0& &h^{21}+1& & h^{21}+1& &0& & \\
% &1& &h^{31}+h^{21}& &h^{22}+2h^{21}+h^{11}& &h^{31}+h^{21}& &1& \\
%0& &h^{31}+1& &h^{31}+h^{22}+2h^{21}& &h^{31}+h^{22}+2h^{21}& &
%h^{31}+1 & &0
%\end{array}
%}\nonumber
%\eeqa 
%and for $CY^3\x K3$
%\beqa
%{\scriptsize
%\begin{array}{ccccccccccc}
% & & & & &1& & & & & \\
% & & & &0& &0& & & & \\
% & & &1& &h^{11}+20& &1& & & \\
% & &1& &h^{21}& &h^{21}& &1& & \\
% &0& &h^{11}& &20h^{11}+1& &h^{11}& &0& \\
%1& &h^{21}+20& &1+21h^{21}& &1+21h^{21}& &h^{21}+20& &1
%\end{array}
%}\nonumber
%\eeqa 

\end{document}